\documentclass{svmult}
\usepackage{graphicx}
\usepackage{cite}
\usepackage{amsmath,amssymb}

\begin{document}

\title*{Astrophysical Axion Bounds}
\label{chap:raffelt-Astrophysical Axion Bounds}

\author{Georg G.~Raffelt}
\titlerunning{Astrophysical Axion Bounds}
\toctitle{Astrophysical Axion Bounds}

\institute{Max-Planck-Institut f\"ur Physik
(Werner-Heisenberg-Institut)\\
  F\"ohringer Ring 6, 80805 M\"unchen, Germany\\
  \texttt{raffelt@mppmu.mpg.de}}

\maketitle

\begin{abstract}
Axion emission by hot and dense plasmas is a new energy-loss channel
for stars. Observational consequences include a modification of the
solar sound-speed profile, an increase of the solar neutrino flux, a
reduction of the helium-burning lifetime of globular-cluster stars,
accelerated white-dwarf cooling, and a reduction of the supernova
SN~1987A neutrino burst duration. We review and update these arguments
and summarize the resulting axion constraints.
\end{abstract}

\section{Introduction}
\label{sec:raffelt-introduction}
The ``outer space'' of astrophysics and cosmology provides a natural
laboratory for the ``inner space'' of elementary particle physics.
Usually one may first think of the early universe or perhaps
high-energy cosmic rays for arguments in favor or against a new
particle-physics model. However, in the case of axions the low
energies available in stars are well suited for very
sensitive~tests.

The basic idea is very simple. Stars are powerful sources for weakly
interacting particles such as neutrinos, gravitons, hypothetical
axions, and other new particles that can be produced by nuclear
reactions or by thermal processes in the stellar interior. Even when
this particle flux cannot be directly measured, the properties of
stars themselves would change if they lost too much energy into a
new channel.  This ``energy-loss argument'' has been widely used to
constrain a long list of particle properties,
see~\cite{GR:Gamow:1940, GR:Gamow:1941, GR:Bernstein:1963qh,
  GR:Stothers:1970ap, GR:Sato:1975vy, GR:Dicus:fp, GR:Vysotsky:1978dc} for
early examples and~\cite{GR:Turner:1989vc, GR:Raffelt:1990yz,
  GR:Raffelt:1996wa, GR:Raffelt:1999tx, GR:Yao:2006px} for extensive
reviews. We summarize here the main arguments that have been put forward,
the observational evidence, and the resulting constraints for ``invisible
axions.''

To this end we review in Sect.~\ref{sec:raffelt-interactions} the
axion interactions with photons and fermions. In
Sect.~\ref{sec:raffelt-sun-as-axion-source} we consider the Sun as
an axion source, notably by the Primakoff process, and review limits
on the axion-photon interaction strength by helioseismology, the
measured neutrino flux, and the CAST experiment. In
Sects.~\ref{sec:raffelt-globularclusters},
\ref{sec:raffelt-whitedwarfs}, and~\ref{sec:raffelt-SN1987A} we
review axion limits from globular cluster stars, white-dwarf
cooling, and supernova SN~1987A, respectively. We summarize these
constraints in Sect.~\ref{sec:raffelt-conclusions} in juxtaposition
with cosmological arguments.

\section{Axion Interactions}
\label{sec:raffelt-interactions}

The particle-physics motivation for ``invisible'' axions and their
main properties were introduced in
Chap.~1 of this volume. Before turning
to their role in stars, we briefly review the phenomenological
properties of these pseudo Nambu-Goldstone bosons of the Peccei
Quinn (PQ) symmetry. The mass and interaction strength with ordinary
particles is approximately given in terms of the relevant $\pi^0$
properties, scaled with $f_\pi/f_{a}$ where $f_\pi=92\,{\rm MeV}$ is
the pion decay constant and $f_{a}$ is the PQ scale or axion decay
constant. The normalization of $f_{a}$ is defined by the axion-gluon
interaction
\begin{equation}
  L_{a\gamma\gamma}=\frac{g_{\rm s}^2}{32\,\pi^2}\,\frac{a}{f_{a}}\,
  G_{\mu\nu}^b\tilde G^{b\mu\nu},
\end{equation}
where $a$ is the axion field, $G$ the gluon field-strength tensor,
$\tilde G$ its dual, and $b$ a color index. Color anomaly factors
have been absorbed in this definition of $f_{a}$, which is the
quantity that is relevant for all low-energy
phenomena~\cite{GR:Georgi:1986df}.

The PQ symmetry is explicitly broken at low energies and axions
acquire a small mass. Unless there are non-QCD contributions,
perhaps from Planck-scale physics~\cite{GR:Kamionkowski:1992mf,
GR:Barr:1992qq}, the mass is
\begin{equation}\label{eq:raffelt-axmass}
  m_{a}=\frac{z^{1/2}}{1+z}\,\frac{f_\pi m_\pi}{f_{a}}
  =\frac{6.0\,{\rm eV}}{f_{a}/10^6\,{\rm GeV}}\;,
\end{equation}
where $z=m_{u}/m_{d}$ is the up/down quark mass ratio. We will
follow the previous axion literature and usually assume the
canonical value $z=0.56$ \cite{GR:Gasser:1982ap,
GR:Leutwyler:1996qg}, although it could vary in the range
$z=0.3$--$0.6$~\cite{GR:Yao:2006px}.

Another generic property of axions is their two-photon interaction
that plays a key role for most searches,
\begin{equation}
  L_{a\gamma\gamma}=
  \frac{g_{a\gamma\gamma}}{4}\,F_{\mu\nu}\tilde F^{\mu\nu}a
  =-g_{a\gamma\gamma}\,\vec{E}\cdot\vec{B}\,a\;.
\end{equation}
Here, $F$ is the electromagnetic field-strength tensor, $\tilde F$
its dual, and $\vec{E}$ and $\vec{B}$ the electric and magnetic
fields, respectively. The coupling constant is
\begin{equation}\label{eq:raffelt-axionphoton}
  g_{a\gamma\gamma}=\frac{\alpha}{2\pi f_{a}}
  \left(\frac{E}{N}-\frac{2}{3}\,\frac{4+z}{1+z}\right)
  =\frac{\alpha}{2\pi}
  \left(\frac{E}{N}-\frac{2}{3}\,\frac{4+z}{1+z}\right)
  \frac{1+z}{z^{1/2}}\,\frac{m_{a}}{m_\pi f_\pi}\;,
\end{equation}
where $E$ and $N$, respectively, are the electromagnetic and color
anomaly of the axial current associated with the axion field.
$E/N=8/3$ in grand unified models, e.g.\ the DFSZ
model~\cite{GR:Zhitnitsky:1980tq, GR:Dine:1981rt}, whereas $E/N=0$
in the KSVZ model~\cite{GR:Kim:1979if, GR:Shifman:1979if}. While
these cases are often used as generic examples, in general $E/N$ is
not known so that for fixed $f_{a}$ a broad range of
$g_{a\gamma\gamma}$ values is possible~\cite{GR:Cheng:1995fd}.
Still, barring fine-tuned cancelations, $g_{a\gamma\gamma}$ scales
from the corresponding pion interaction by virtue of the
relation~(\ref{eq:raffelt-axionphoton}). Taking the model-dependent
factors to be of order unity, this relation defines the ``axion
line'' in the $m_{a}$--$g_{a\gamma\gamma}$ plane.

Axions or axion-like particles with a two-photon vertex decay into
two photons with a rate
\begin{eqnarray}
  \Gamma_{{a}\to\gamma\gamma}&=&\frac{g_{a\gamma\gamma}^2m_{a}^3}{64\,\pi}
  =\frac{\alpha^2}{256\,\pi^3}
  \left[\left(\frac{E}{N}-\frac{2}{3}\,\frac{4+z}{1+z}\right)
    \frac{1+z}{z^{1/2}}\right]^2\,\frac{m_{a}^5}{m_\pi^2 f_\pi^2}
  \nonumber\\
  &=&1.1\times10^{-24}\,{\rm s}^{-1}\,
  \left(\frac{m_{a}}{\rm eV}\right)^5,
\end{eqnarray}
where the first expression is for general pseudoscalars, the second
applies specifically to axions, and the numerical one assumes
$z=0.56$ and the hadronic case $E/N=0$. Comparing with the age of
the universe of $4.3\times10^{17}\,{\rm s}$ reveals that axions
decay on a cosmic time scale if $m_{a}\gtrsim20\,{\rm eV}$.

The interaction with fermions $j$ has a derivative structure so that
it is invariant under $a\to a+a_0$ as behooves a Nambu-Goldstone
boson,
\begin{equation}\label{eq:raffelt-axionfermioncoupling}
  {L}_{ajj}=\frac{C_{j}}{2f_{a}}\,
  \bar\Psi_j\gamma^\mu\gamma_5\Psi_j\partial_\mu a
  \hbox{\quad or\quad}
  -\I\,\frac{C_{j} m_j}{f_{a}}\,\bar\Psi_j\gamma_5\Psi_ja\;.
\end{equation}
Here, $\Psi_j$ is the fermion field, $m_j$ its mass, and $C_{j}$ a
model-dependent numerical coefficient. The combination
$g_{ajj}\equiv C_{j} m_j/f_{a}$ plays the role of a Yukawa coupling
and $\alpha_{ajj}\equiv g_{ajj}^2/4\pi$ of a ``fine-structure
constant.''  The pseudoscalar form is usually equivalent to the
derivative structure, but one has to be careful in processes where
two Nambu-Goldstone bosons are attached to one fermion line, for
example an axion and a pion attached to a nucleon in the context of
axion emission by nucleon bremsstrahlung~\cite{GR:Raffelt:1987yt,
GR:Carena:1988kr}.

In hadronic models such as KSVZ~\cite{GR:Kim:1979if,
GR:Shifman:1979if}, axions do not couple to ordinary quarks and
leptons at tree level, whereas in the DFSZ
model~\cite{GR:Kim:1979if, GR:Shifman:1979if}
\begin{equation}
  C_e=\frac{\cos^2\beta}{3}\;.
\end{equation}
Here, $\cot\beta$ is the ratio of two Higgs vacuum expectation
values of this model.

For nucleons, the dimensionless couplings $C_{n,p}$ are related by
generalized Goldberger-Treiman relations to nucleon axial-vector
current matrix elements,
\begin{eqnarray}
  C_p&=&\left(C_u-\eta\right)\Delta u
       +\left(C_d-\eta z\right)\Delta d
       +\left(C_s-\eta w\right)\Delta s\;,\nonumber\\
 C_n&=&\left(C_u-\eta\right)\Delta d
       +\left(C_d-\eta z\right)\Delta u
       +\left(C_s-\eta w\right)\Delta s\;.
\end{eqnarray}
Here, $\eta=(1+z+w)^{-1}$ with $z=m_u/m_d$ and $w=m_u/m_s\ll z$. The
quantities $\Delta q$ represent the axial-vector current couplings
to the proton, $\Delta q\, S_\mu=\langle p|\bar q\gamma_\mu\gamma_5
q|p\rangle$ where $S_\mu$ is the proton spin.

Neutron beta decay and strong isospin symmetry tell us that $\Delta
u-\Delta d=F+D=1.267\pm0.0035$ whereas hyperon decays and flavour
SU(3) symmetry imply $\Delta u+\Delta d-2\Delta
s=3F-D=0.585\pm0.025$. Recent determinations of the strange-quark
contribution are $\Delta s=-0.08\pm0.01_{\rm stat}\pm0.05_{\rm
syst}$ from the COMPASS experiment~\cite{GR:Alexakhin:2006vx} and
$\Delta s=-0.085\pm0.008_{\rm
  exp}\pm0.013_{\rm theor}\pm0.009_{\rm evol}$ from
HERMES~\cite{GR:Airapetian:2006vy}, in agreement with each other and with
an early estimate of $\Delta s=-0.11\pm0.03$ \cite{GR:Ellis:1995de}. We
thus adopt the estimates
\begin{eqnarray}
  \Delta u&=&+0.841\pm0.020\;,\nonumber\\
  \Delta d&=&-0.426\pm0.020\;,\nonumber\\
  \Delta s&=&-0.085\pm0.015\;,
\end{eqnarray}
that are very similar to previous values used in the axion literature.

The uncertainty of the axion-nucleon couplings is dominated by the
large uncertainty of $z=0.3$--0.6 that was mentioned above. For
hadronic axions we have $C_{u,d,s}=0$ so that $C_p=-0.55$ and
$C_n=+0.14$ for $z=0.3$ and $C_p=-0.37$ and $C_n=-0.05$ for $z=0.6$.
Therefore, while it is well possible that $C_n=0$, $C_p$ does not
vanish within the plausible $z$ range. In the DFSZ model we have
$C_u=\frac{1}{3}\sin^2\beta$ and $C_d=\frac{1}{3}\cos^2\beta$. Even
with the large allowed $z$ range, $C_n$ and $C_p$ never vanish
simultaneously. An extreme case is $\cos^2\beta=0$ where $C_p=0$ for
$z=0.3$, but in this case $C_n=-0.27$.

\section{The Sun as an Axion Source}
\label{sec:raffelt-sun-as-axion-source}

\subsection{Axion Flux from the Primakoff Process}
\label{subsec:raffelt-axion-flux-from}

The Sun would be a powerful axion source. This flux can be searched
directly, notably by the CAST experiment. Its sensitivity is
competitive with the globular cluster limits (Sect.
\ref{sec:raffelt-globularclusters}) for hadronic models. In this
case the dominant emission process is the Primakoff
effect~\cite{GR:Primakoff}, i.e., particles with a two-photon vertex
transform into photons in external electric or magnetic fields.
Therefore, stars produce axions from thermal photons in the
fluctuating electromagnetic fields of the stellar
plasma~\cite{GR:Dicus:fp}.

Calculating the solar axion flux is straightforward except for the
proper inclusion of screening effects~\cite{GR:Raffelt:1985nk,
GR:Altherr:1993zd}. The transition rate for a photon of energy $E$
into an axion of the same energy (recoil effects are neglected)
is~\cite{GR:Raffelt:1987np}
\begin{equation}\label{eq:raffelt-Gamma-ag}
  \Gamma_{\gamma\to {a}}= \frac{g_{a\gamma\gamma}^2T\kappa_{\rm s}^2}{32
    \pi} \bigg[\bigg(1+\frac{\kappa_{\rm s}^2}{4E^2}\bigg)
  \ln\bigg(1+\frac{4E^2}{\kappa_{\rm s}^2}\bigg)-1\bigg]\;,
\end{equation}
where $T$ is the temperature (natural units with $\hbar=c=k_{\rm
B}=1$ are used). The screening scale in the Debye-H\"uckel
approximation~is
\begin{equation}
  \kappa_{\rm s}^2= \frac{4\pi\alpha}{T}\biggl(n_{e}+\sum_{\rm nuclei}
  Z_j^2n_j\biggr)\;,
\end{equation}
where $n_{e}$ is the electron density and $n_j$ that of the $j$-th
ion of charge $Z_j$. Near the solar center $\kappa_{\rm s}\approx
9\,{\rm keV}$. Note that $(\kappa_{\rm s}/T)^2\approx12$ is nearly
constant throughout the Sun whereas it is about 2.5 throughout the
core of a low-mass helium-burning star.

Ignoring the plasma frequency for the initial-state photons, the
energy-loss rate per unit volume is~\cite{GR:Raffelt:1996wa,
GR:Raffelt:1987np}
\begin{equation}\label{eq:raffelt-primakofflossrate}
  Q=\frac{g_{a\gamma\gamma}^2T^7}{4\pi}\,F\;,
\end{equation}
where $F$ is a numerical factor of order unity. For $(\kappa_{\rm
  s}/T)^2=2.5$ and 12 one finds $F=0.98$ and $1.84$, respectively.

Integrating over a standard solar model, one finds an axion flux at
Earth that is is well approximated by ($E$ in keV)
\begin{equation}
  \label{eq:raffelt-bestfit}
  \frac{\D\Phi_{a}}{\D E}=g_{10}^2\,\,6.0\times
  10^{10}\,{\rm cm}^{-2}\,{\rm s}^{-1}\,{\rm keV}^{-1}\,
  E^{2.481}\, {\rm e}^{-E/1.205}\;,
\end{equation}
where $g_{10}=g_{a\gamma\gamma}/(10^{-10}\,{\rm GeV}^{-1})$. The
integrated flux parameters are
\begin{eqnarray}
  \Phi_{a}&=&g_{10}^2\,3.75\times10^{11}\,{\rm cm}^{-2}\,{\rm s}^{-1}\;,
  \nonumber\\
  L_{a}&=&g_{10}^2\,1.85\times 10^{-3} L_\odot\;.
\end{eqnarray}
The maximum of the distribution is at $3.0\,{\rm keV}$, the average
energy is $4.2\,{\rm keV}$.

\subsection{Solar Age}
\label{subsec:raffelt-solar-age}

The properties of the Sun itself constrain this flux. The axion
losses lead to an enhanced consumption of nuclear fuel. The standard
Sun is halfway through its hydrogen-burning phase so that the solar
axion luminosity should not exceed its photon luminosity $L_\odot$.

As an example we recall that a magnetically induced vacuum dichroism
observed by the PVLAS experiment~\cite{GR:Zavattini:2005tm}, if
interpreted in terms of an axion-like particle (ALP),
requires~\cite{GR:Zavattini:2005tm, GR:Cameron:1993mr}
$g_{a\gamma\gamma}=\hbox{2--5}\times10^{-6}\,{\rm GeV}^{-1}$ and
$m_{a}=\hbox{1--1.5}\,{\rm meV}$.  With this coupling strength, the
Sun's ALP luminosity would exceed $L_\odot$ by a factor of a million
and thus could live only for about 1000~years. Perhaps this problem
can be circumvented, but it is noteworthy that even a crude
astrophysical argument severely constrains the particle
interpretation of the PVLAS signature.

\subsection{Helioseismology}
\label{subsec:raffelt-helioseismology}

For a more refined constraint we note that a model of the
present-day Sun, with the integrated effect of axion losses taken
into account, would differ from a standard solar model. The modified
sound-speed profile can be diagnosed by helioseismology, providing a
conservative limit~\cite{GR:Schlattl:1998fz}
\begin{equation}\label{eq:raffelt-helioseismology}
  g_{a\gamma\gamma}\lesssim 1\times10^{-9}\,{\rm GeV}^{-1}\;,
\end{equation}
corresponding to $L_{a}\lesssim 0.20\,L_\odot$. More recent
determinations of the solar metal abundances have diminished the
agreement between standard solar models and
helioseismology~\cite{GR:Bahcall:2004pz}, but these modifications do
not change the limit~(\ref{eq:raffelt-helioseismology}).

\subsection{Solar Neutrino Flux}
\label{subsec:solar-neutrino-flux}

The energy loss by solar axion emission requires enhanced nuclear
burning and thus a somewhat increased temperature in the Sun.
Self-consistent solar models with axion losses reveal that
$g_{a\gamma\gamma}=4.5\times10^{-10}\,{\rm GeV}^{-1}$ implies a 20\%
increase of the solar $^8$B neutrino flux~\cite{GR:Schlattl:1998fz}.
For $g_{a\gamma\gamma}=10\times10^{-10}\,{\rm GeV}^{-1}$ the
increase would be a factor of~2.4.

The measured all-flavor $^8$B neutrino flux is $4.94\times10^6\,{\rm
cm^{-2}\,s^{-1}}$ with an uncertainty of about 8.8\%
\cite{GR:Ahmad:2002jz, GR:Aharmim:2005gt}. The old standard solar
model predictions were about 5.7--5.9 in the same units, whereas the
new metal abundances imply 4.5--4.6, each time with a 16\%
``theoretical $1\sigma$ error'' \cite{GR:Bahcall:2004pz}. Therefore,
the measured neutrino fluxes imply a limit
\begin{equation}\label{eq:raffelt-solarneutrinos}
  g_{a\gamma\gamma}\lesssim 5\times10^{-10}\,{\rm GeV}^{-1}\;,
\end{equation}
corresponding to $L_{a}\lesssim0.04\,L_\odot$. A more precise limit
with a realistic error budget would require self-consistent solar
models on a finer spacing of $g_{a\gamma\gamma}$.

\subsection{Searches for Solar Axions}
\label{subsec:raffelt-search-solar-axions}

The solar axion flux can be searched with the inverse Primakoff
process where axions convert to photons in a macroscopic $B$ field,
the ``axion helioscope'' technique~\cite{GR:Sikivie:1983ip}. One
would look at the Sun through a ``magnetic telescope'' and place an
x-ray detector at the far end. The conversion can be coherent over a
large propagation distance and is then pictured as a particle
oscillation effect~\cite{GR:Raffelt:1987im}.

Early helioscope searches were performed in
Brookhaven~\cite{GR:Lazarus:1992ry} and Tokyo
\cite{GR:Moriyama:1998kd, GR:Inoue:2002qy}. Solar axions could also
transform in electric crystal fields, but the limits obtained by
SOLAX~\cite{GR:Avignone:1997th}, COSME~\cite{GR:Morales:2001we}, and
DAMA~\cite{GR:Bernabei:ny} are less restrictive and require a solar
axion luminosity exceeding~(\ref{eq:raffelt-helioseismology})
and~(\ref{eq:raffelt-solarneutrinos}), i.e., these limits are not
self-consistent.

The first helioscope that can actually reach the ``axion line'' is
the CERN Axion Solar Telescope (CAST). The non-observation of a
signal above background leads to a
constraint~\cite{GR:Andriamonje:2004hi}
\begin{equation}\label{eq:raffelt-castlimit}
  g_{a\gamma\gamma}<1.16\times 10^{-10}\,{\rm GeV}^{-1}
  \hbox{\quad(95\% CL)\qquad for\quad}m_{a}\lesssim0.02\,{\rm eV}
  \;.
\end{equation}
For larger masses, the axion-photon transition is suppressed by the
energy-momentum mismatch between particles of different mass. The
full rate can be restored in a narrow range of masses by providing
the photons with a refractive mass in the presence of a low-$Z$
gas~\cite{GR:vanBibber:1988ge}, a method that was already used in
the Tokyo experiment~\cite{GR:Inoue:2002qy} and is also used in the
ongoing CAST Phase~II. CAST is foreseen to reach eventually
$m_{a}\lesssim 1\,{\rm eV}$.

\subsection{Do Axions Escape from the Sun?}
\label{subsec:raffelt-do-axions-escape}

CAST can detect axions only if they actually escape from the Sun.
Their mean free path (mfp) against the Primakoff process is the
inverse of~(\ref{eq:raffelt-Gamma-ag}). For $4\,{\rm keV}$ axions
and with $T\approx 1.3\,{\rm keV}$ and $\kappa_{\rm s}\approx9\,{\rm
keV}$ at the solar center we find $\lambda_{a}\approx
g_{10}^{-2}\,6\times10^{24}\,{\rm cm}\approx
g_{10}^{-2}\,8\times10^{13}\,R_{\odot}$, or about $10^{-3}$ of the
radius of the visible universe. Therefore, $g_{a\gamma\gamma}$ would
have to be more than $10^7$ larger than the CAST limit for axions to
be re-absorbed in the Sun.

Even in this extreme case they are not harmless because they would
carry the bulk of the energy flux that otherwise is carried by
photons. The mfp of low-mass particles in the trapping regime should
be shorter than that of photons (about $10\,{\rm cm}$ near the solar
center) to avoid a dramatic modification of the solar
structure~\cite{GR:Raffelt:1988rx}. This requirement is so extreme
that for anything similar to axions the possibility of re-absorption
is not a serious possibility.

\section{Globular-Cluster Stars}
\label{sec:raffelt-globularclusters}

\subsection{Helium-Burning Lifetime and the Axion-Photon
  Interaction}
\label{subsec:raffelt-helium-burn-lifet}

\begin{figure}[t]
  \centering
  \includegraphics[width=0.70\textwidth]{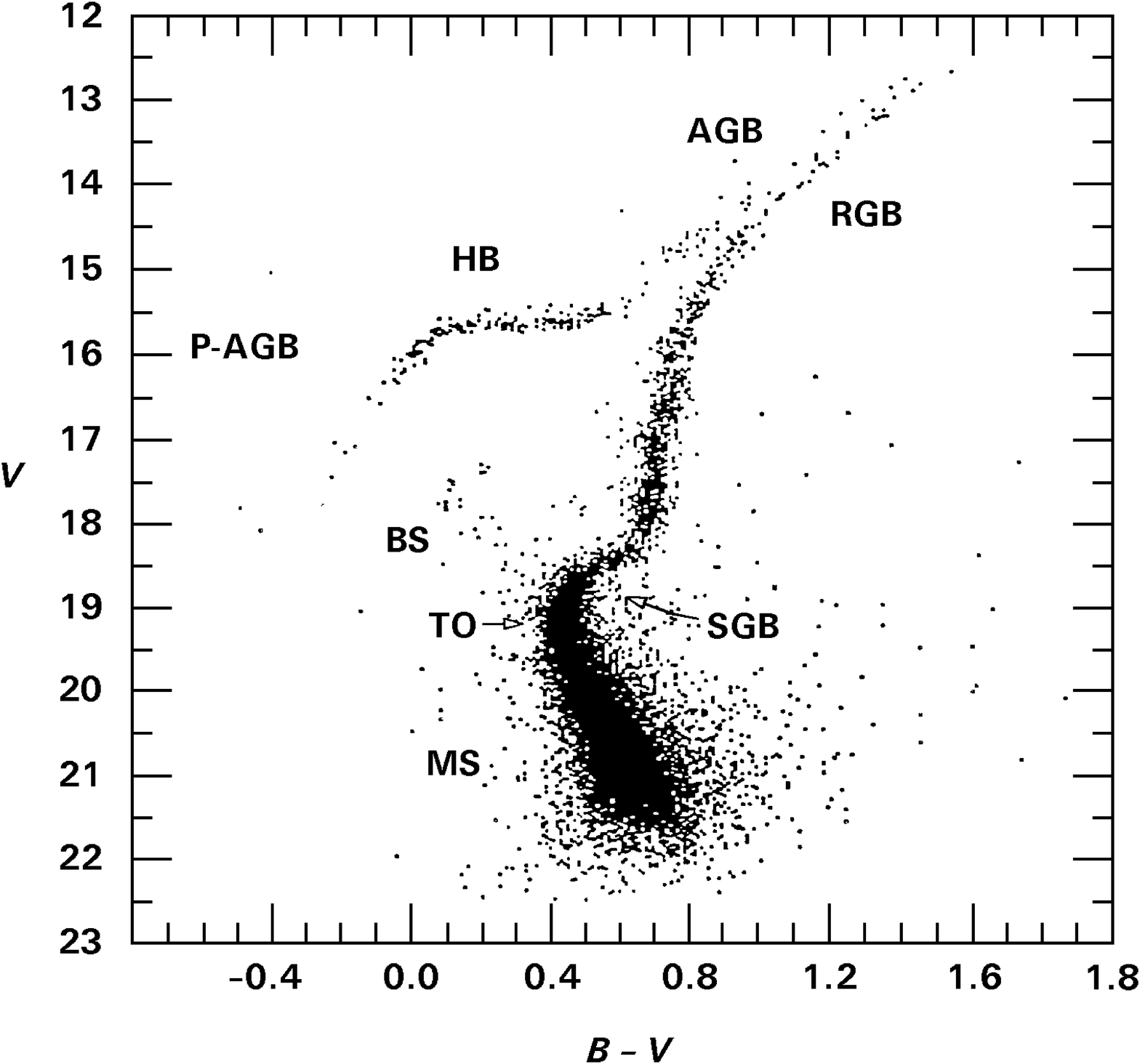}
  \caption{Color-magnitude diagram for the globular cluster M3, based
    on 10,637 stars~\cite{GR:Buonanno:1986}. Vertically is the brightness
    in the visual (V) band, horizontally the difference between B (blue)
    and V brightness, i.e.\ a measure of the color and thus surface
    temperature, where blue (hot) stars lie toward the left.  The
    classification for the evolutionary phases is as
    follows~\cite{GR:Renzini:1988}.  MS (main sequence): core hydrogen
    burning. BS (blue stragglers). TO (main-sequence turnoff): central
    hydrogen is exhausted.  SGB (subgiant branch): hydrogen burning in a
    thick shell. RGB (red-giant branch): hydrogen burning in a thin shell
    with a growing core until helium ignites. HB (horizontal branch):
    helium burning in the core and hydrogen burning in a shell. AGB
    (asymptotic giant branch): helium and hydrogen shell burning. P-AGB
    (post-asymptotic giant branch): final evolution from the AGB to the
    white-dwarf stage \label{fig:raffelt-colmag}}
\end{figure}

A restrictive limit on $g_{a\gamma\gamma}$ arises from
globular-cluster stars. A globular cluster is a gravitationally
bound system of stars that formed at the same time and thus differ
primarily in their mass. A globular cluster provides a homogeneous
population of stars, allowing for detailed tests of
stellar-evolution theory. The stars surviving since formation have
masses somewhat below $1\,M_\odot$.  In a color-magnitude diagram
(Fig. \ref{fig:raffelt-colmag}), where one plots essentially the
surface brightness vs.\ the surface temperature, stars appear in
characteristic loci, allowing one to identify their state of
evolution.

The stars on the horizontal branch (HB) have reached helium burning,
where their core (about $0.5\,M_\odot$) generates energy by fusing
helium to carbon and oxygen with a core-averaged energy release of
about $80\,\rm erg\,g^{-1}\,s^{-1}$. A typical density is
$10^4\,{\rm g}\,{\rm cm}^{-3}$ and a typical temperature $10^8\,{\rm
K}$. The Primakoff energy loss
rate~(\ref{eq:raffelt-primakofflossrate}) implies that the
energy-loss rate per unit mass, $\varepsilon=Q/\varrho$, is
proportional to $T^7/\varrho$. Averaged over a typical HB-star core
one finds $\langle (T/10^8\,{\rm
  K})^7\,(10^4\,{\rm g}\,{\rm cm}^{-3}/\varrho)\rangle\approx0.3$.
Therfore, the core-averaged energy loss rate is about $g_{10}^2\,30\,\rm
erg\,g^{-1}\,s^{-1}$. The main effect would be accelerated consumption of
helium and thus a reduction of the HB lifetime by a factor
$80/(80+30\,g_{10}^2)$, i.e., by about 30\% for $g_{10}=1$.

The HB lifetime can be measured relative to the red-giant
evolutionary time scale by comparing the number of HB stars with the
number of RGB stars that are brighter than the HB. Number counts in
15~globular clusters~\cite{GR:Buzzoni:1983} show that this number
ratio agrees with expectations within 20--40\% in any one cluster,
where the error is mostly statistical because typically only about
100 HB stars were present in the used fields of view. Compounding
the results of all 15 clusters, the helium-burning lifetime agrees
with expectations to within about 10\% \cite{GR:Raffelt:1996wa,
GR:Raffelt:1999tx}. Of course, with modern data these results likely
could be improved. Either way, a reasonably conservative limit~is
\begin{equation}
  g_{a\gamma\gamma}<10^{-10}\,{\rm GeV}^{-1}\;.
\end{equation}
It is comparable to the CAST limit~(\ref{eq:raffelt-castlimit}), but
applies for higher masses. The relevant temperature is about
$10\,{\rm
  keV}$ so that significant threshold effects begin only at about
$m_{a}\gtrsim 30\,{\rm keV}$.  For QCD axions the coupling increases
with mass so that the limit reaches to even larger masses.

In the helium-burning core, convection and semi-convection dredges
helium to the burning site so that 25--30\% of all helium is burnt
during the HB phase. Therefore, while the standard theoretical
predictions depend on a phenomenological treatment of convection,
there is limited room for additional energy supply, even if the
treatment of convection were grossly incorrect.

\subsection{Helium Ignition and the Axion-Electron Interaction}
\label{subsec:raffelt-helium-ignition-axion}

Stars on the red-giant branch (RGB) have a degenerate helium core
with a typical density $10^6\,{\rm g}\,{\rm cm}^{-3}$ and $T\approx
10^8\,{\rm
  K}$. Helium ignites at a critical combination of $\varrho$ and $T$.
Therefore, helium ignition can be delayed by axion cooling.  This implies
that the core grows more massive before helium ignites.  One consequence is
that the RGB will extend to brighter stars, i.e., the brightness of the
brightest red giant in a given globular cluster signifies the core mass at
helium ignition.  Detailed studies reveal that the core mass at helium
ignition agrees with theoretical expectations within 5--10\%
\cite{GR:Raffelt:1996wa, GR:Raffelt:1989xu, GR:Raffelt:1990pj,
  GR:Raffelt:1994ry, GR:Catelan:1995ba}. In turn, this implies that a novel
energy-loss rate at $T=10^8\,{\rm K}$ and an average density
$\langle\varrho\rangle=2\times10^5\,{\rm g}\,{\rm cm}^{-3}$ should not
exceed about $10\,{\rm erg}\,{\rm g}^{-1}\,{\rm s}^{-1}$. At these
conditions the standard neutrino emission is about $4\,{\rm erg}\,{\rm
  g}^{-1}\,{\rm s}^{-1}$.

The helium-burning lifetime is useful to constrain the axion-photon
interaction because the Primakoff rate is suppressed in the
degenerate red-giant cores and thus is more effective in HB stars.
The helium-ignition argument, on the other hand, is useful when the
emission rate is larger on the RGB than on the HB as for
bremsstrahlung $e+Ze\to Ze+e+a$. For the conditions in a red-giant
core one finds $\varepsilon_{\rm
  brems}\approx\alpha_{aee}\,2\times10^{27}\,{\rm erg}\,{\rm g}^{-1}\,{\rm
  s}^{-1}$ \cite{GR:Raffelt:1994ry} so that
\begin{equation}\label{eq:raffelt-ignitionlimit}
  \alpha_{aee}<0.5\times10^{-26}
  \hbox{\quad or\quad}
  g_{aee}<3\times10^{-13}\;.
\end{equation}
In the DFSZ model this limit corresponds to
$f_{a}/\cos^2\beta>0.8\times10^9\,{\rm GeV}$, $m_{a}<9\,{\rm
meV}/\cos^2\beta$ and
$g_{a\gamma\gamma}\cos^2\beta<1.2\times10^{-12}\,{\rm GeV}^{-1}$.

\subsection{Asymptotic Giant Branch (AGB) Evolution}
\label{subsec:raffelt-asymptotic-giant-branch}

For axion-electron interactions near or even below the
bound~(\ref{eq:raffelt-ignitionlimit}), the emission will strongly
affect the evolutionary behavior of AGB
stars~\cite{GR:Dominguez:1999gg}. However, these results have not
been linked closely enough to observational data to obtain new
limits or discover evidence for axion emission.

\section{White-Dwarf Cooling}
\label{sec:raffelt-whitedwarfs}

The degenerate core of a low-mass red giant before helium ignition
is essentially a helium white dwarf. After the HB phase, when helium
burning has ended, low-mass stars once more ascend the red-giant
branch as ``asymptotic giants'' (AGB stars). They have a degenerate
carbon-oxygen core and helium burning in a shell. Fast mass loss
creates a ``planetary nebula'' surrounding a compact remnant, a
white dwarf, that first cools by neutrino emission and later by
surface photon emission.

The observed white-dwarf luminosity function reveals that their
cooling speed agrees with expectations, constraining new cooling
agents such as axion emission~\cite{GR:Raffelt:1996wa,
GR:Raffelt:1985nj, GR:Wang:1992gc, GR:Blinnikov:1994}. The resulting
limit on the axion-electron coupling of $\alpha_{aee}\lesssim
1\times10^{-26}$ is comparable to the globular-cluster limit
of~(\ref{eq:raffelt-ignitionlimit}).

The cooling speed of individual white dwarfs can be estimated in
some cases where they appear as ZZ Ceti stars, i.e., when they are
pulsationally unstable and when the period decrease $\dot P/P$ can
be measured, a quantity that is sensitive to the cooling speed. A
well-studied case is the star G117--B15A. For some time it seemed to
be cooling too fast, an effect that could have been attributed to
axion cooling with $\alpha_{aee}=0.2$--$0.8\times10^{-26}$
\cite{GR:Isern:1992}. More recent analyses no longer require a new
cooling channel, allowing one to set a limit on axion losses
corresponding to\footnote{The limit on $g_{aee}$
  stated in~\cite{GR:Corsico:2001be} is an order of magnitude more
  restrictive, but this is an obvious misprint. Likewise, their stated
  limit $m_{a}<5\,{\rm meV}\,\cos^2\beta$ has an incorrect scaling with
  $\cos^2\beta$.} \cite{GR:Corsico:2001be, GR:Isern:2003xj}
\begin{equation}
  \alpha_{aee}<1.3\times10^{-27}
  \hbox{\quad or\quad}
  g_{aee}<1.3\times10^{-13}
\end{equation}
at a statistical 95\% CL. In the DFSZ model this implies
$m_{a}<5\,{\rm meV}/\cos^2\beta$. This is the most restrictive limit
on the axion-electron interaction.

\section{Supernova 1987A}
\label{sec:raffelt-SN1987A}

\subsection{Energy-Loss Argument}
\label{subsec:raffelt-energy-loss-argument}

About two dozen neutrinos from SN~1987A were observed almost twenty
years ago in several detectors~\cite{GR:Koshiba:1992yb}.  The total
number of events, their energies, and the distribution over several
seconds correspond reasonably well to theoretical expectations. In
the standard picture~\cite{GR:Burrows:2000mk, GR:Woosley:2006ie},
the core collapse of a massive star leads to a proto neutron star, a
solar-mass object at nuclear density and a temperature of several
$10\,{\rm MeV}$, where even neutrinos are trapped. The long time
scale of emission is explained by diffusive neutrino energy
transport. The emission of more weakly interacting particles can be
a more efficient energy-loss channel, resulting in a reduced
neutrino burst duration. The late-time signal is most sensitive to
such losses because the early neutrino emission is powered by
accretion and thus not very sensitive to volume losses.

\begin{figure}
  \centering
  \includegraphics[width=0.60\textwidth]{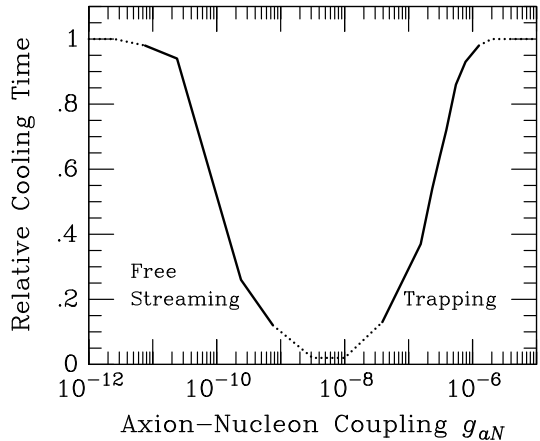}
  \caption{Relative duration of a SN neutrino burst as a function of
    the axion-nucleon coupling~\cite{GR:Raffelt:1996wa}. Freely streaming
    axions are emitted from the entire core volume, trapped ones from an
    ``axion sphere.'' The solid line is from numerical
    calculations~\cite{GR:Burrows:1988ah, GR:Burrows:1990pk}.  The dotted
    line is an arbitrary continuation to guide the
    eye\label{fig:raffelt-pulse}}
\end{figure}

This argument has been applied to many cases, from right-handed
neutrinos to Kaluza-Klein gravitons, but axions are the earliest and
most widely discussed example~\cite{GR:Raffelt:1987yt,
GR:Ellis:1987pk, GR:Turner:1987by, GR:Mayle:1987as, GR:Mayle:1989yx,
GR:Brinkmann:1988vi, GR:Burrows:1988ah, GR:Burrows:1990pk,
GR:Janka:1995ir, GR:Keil:1996ju, GR:Hanhart:2000ae}. They are
emitted by nucleon bremsstrahlung $N+N\to N+N+a$ that depends on the
axion-nucleon Yukawa coupling $g_{aNN}$, here taken to be an average
of the couplings to neutrons and protons. Figure
\ref{fig:raffelt-pulse} illustrates that axion emission leaves the
signal duration unchanged when $g_{aNN}$ is very small.  For larger
couplings, the signal is shortened until it reaches a minimum,
roughly when the axion mfp corresponds to the geometric size of the
SN core. For even larger couplings, axions are trapped and are
emitted from an ``axion sphere.'' When it moves beyond the neutrino
sphere, the signal duration once more remains unaffected.

Of course, such ``strongly'' interacting axions are not necessarily
harmless. They may play an important role during the infall phase.
Moreover, in the water Cherenkov detectors that registered the
SN~1987A neutrinos, these axions would have interacted with oxygen
nuclei, leading to the release of $\gamma$ rays and causing too many
events~\cite{GR:Engel:1990zd}.

However, for axions and other particles, the trapping regime is
usually excluded by other arguments so that the free-streaming
regime is of greater interest. An approximate analytic constraint on
the energy-loss rate is~\cite{GR:Raffelt:1990yz}
\begin{equation}\label{eq:raffelt-SN1987A}
  \varepsilon_a\lesssim1\times10^{19}\,{\rm erg\,g^{-1}\,s^{-1}}\;,
\end{equation}
to be calculated at $\varrho=3\times10^{14}\,{\rm g}\,{\rm cm}^{-3}$
and $T=30\,{\rm MeV}$. If we take the SN core to have a mass of
about $1\,M_\odot=2\times10^{33}\,{\rm g}$, this corresponds to an
axion luminosity $L_{a}=\varepsilon_a M_\odot=2\times10^{52}\,{\rm
  erg\,s^{-1}}$.  The gravitational binding energy of the neutron star is
about $3\times10^{53}\,{\rm erg}$ and the emission lasts up to $10\,{\rm
  s}$, i.e., axion losses would compete significantly with neutrino
emission.

We stress that the criterion~(\ref{eq:raffelt-SN1987A}) is not
arbitrary, but was distilled from several numerical simulations that
consistently showed that the burst duration was roughly halved when
the limit~(\ref{eq:raffelt-SN1987A}) was
saturated~\cite{GR:Raffelt:1990yz}. Axion losses are then not small
so that $T=30\,{\rm MeV}$ is not the unperturbed temperature of
these models.  Different numerical models with different input
physics probably have internal temperatures that are more similar
once significant axion losses are included. In any event,
(\ref{eq:raffelt-SN1987A}) represents quite accurately the results
from different simulations.

Recently, self-consistent cooling calculations were performed for
Kaluza-Klein gravitons~\cite{GR:Hanhart:2001fx}, once more
confirming (\ref{eq:raffelt-SN1987A}). The neutrino signal duration
was directly compared with the data and the limit, corresponding
to~(\ref{eq:raffelt-SN1987A}), was found to have a 95\% statistical
CL.

Of course, this and any other numerical study relies on input
physics for which systematic uncertainties are difficult to
quantify, notably the nuclear equation of state and the neutrino
opacities. In addition, the data are very sparse so that any
conclusion based on them suffers from the usual problems of
small-number statistics. Therefore, (\ref{eq:raffelt-SN1987A})
should be viewed as a reasonable guide as to where a new energy-loss
channel causes a significant tension with the SN~1987A pulse
duration.

\subsection{Axion Emission from a Nuclear Medium}
\label{subsec:raffelt-axion-emission-from-a-nuclear-medium}

In order to apply (\ref{eq:raffelt-SN1987A}) to axions one needs the
emission rate from a hot medium at nuclear density. The main
emission process is nucleon bremsstrahlung, \hbox{$N+N\to N+N+a$},
but a reliable calculation of the rate is difficult. Axions couple
to the nucleon spin so that bremsstrahlung requires the spin to
``jiggle'' in a collision, i.e., spin-conserving interactions do not
contribute. This leaves the nuclear tensor force that is only
crudely modeled by one-pion exchange (OPE). In a dense medium, other
problems include the modification of particle masses and couplings
as well as many-body and multiple-scattering~effects.

One approach to estimate the emission rate relies on linear response
theory where emission, absorption, and scattering of neutrinos,
axions and other particles depend only on a few ``form factors'' of
the medium, i.e., the dynamical structure functions
\cite{GR:Janka:1995ir, GR:Raffelt:1991pw, GR:Raffelt:1993ix,
GR:Raffelt:1996za, GR:Raffelt:1996di, GR:Sigl:1995ac,
GR:Raffelt:1998pa, GR:Yamada:1999va, GR:Sedrakian:2000kc,
GR:vanDalen:2003zw}. This approach is perturbative to lowest order
in the weak interaction between neutrinos (or axions) and nucleons,
whereas the interactions among the medium constituents are lumped
into the structure functions.

Assuming the medium to consist of one species of nonrelativistic
nucleons, the relevant quantity is the dynamical spin-density
structure function~\cite{GR:Janka:1995ir}
\begin{equation}\label{eq:raffelt-strucdefinition}
  S_{\sigma}(\omega,{\bf k})=\frac{4}{3n_B}
  \int_{-\infty}^{+\infty}\D t\,{\rm e}^{{\rm i}\omega t}
  \left\langle\hbox{\boldmath$\sigma$}(t,{\bf k})
    \cdot\hbox{\boldmath$\sigma$}(0,-{\bf k})
  \right\rangle\;,
\end{equation}
where $n_B$ is the nucleon (baryon) density and
{\boldmath$\sigma$}$(t,{\bf k})$ the spatial Fourier transform of
the nucleon spin-density operator. The basic principles of quantum
mechanics imply the detailed-balancing condition
\begin{equation}
  S_{\sigma}(-\omega,{\bf k})=S_{\sigma}(\omega,{\bf k})\,
  {\rm e}^{-\omega/T}\;.
\end{equation}
The structure function obeys the sum rule
\begin{equation}\label{eq:raffelt-sumrule}
  \int_{-\infty}^{+\infty}\frac{\D\omega}{2\pi}\,
  S_\sigma(\omega,{\bf k})=
  1+\frac{4}{3n_B}\left\langle
    \sum_{i,j=1\atop i\not=j}^{N_B}{\bf s}_i\cdot {\bf s}_j\,
    \cos({\bf k}\cdot{\bf r}_{ij})\right\rangle\;,
\end{equation}
where ${\bf s}_i$ is the spin operator of nucleon $i$. The f-sum
rule includes a factor $\omega$ under the integral and establishes a
relation to the average nucleon-nucleon spin interaction
energy~\cite{GR:Sigl:1995ac}. It is often assumed that also the
higher sums $\int \D\omega\,\omega^n\,S(\omega,{\bf k})$ exist for
all $n$.

The axion absorption rate and the volume energy loss rate are given
in terms of the structure function as
\begin{eqnarray}
  \Gamma_a&=&\left(\frac{C_N}{2f_{a}}\right)^2\,\frac{n_B}{2}\,
  \omega\,S_\sigma(\omega,k)\;,
  \nonumber\\
  Q_a&=&\left(\frac{C_N}{2f_{a}}\right)^2\frac{n_B}{4\pi^2}
  \int_0^\infty \D\omega\,\omega^4\,S_\sigma(-\omega,k)\;,
\end{eqnarray}
where $k=|{\bf k}|\approx\omega$ is the modulus of the axion
momentum. Neutrino scattering, emission and absorption rates based
on the axial vector current are given by similar phase-space
integrals.

A reliable expression for $S_\sigma(\omega,k)$ is not available so
that we need to use heuristic reasoning. The large nucleon mass
compared with the emitted axion energy suggests to use the
long-wavelength approximation $S_\sigma(\omega)=\lim_{{\bf k}\to
0}S_\sigma(\omega,{\bf k})$, i.e., we neglect the momentum transfer
to the medium. In this limit (\ref{eq:raffelt-strucdefinition})
represents essentially the Fourier transform of the autocorrelation
function of a single nucleon spin.

If we picture the nucleon spin as a classical vector that is kicked
by a random force, we find~\cite{GR:Raffelt:1993ix}
\begin{equation}
 S_\sigma^{\rm class}(\omega)=
 \frac{\Gamma_\sigma}{\omega^2+\Gamma_\sigma^2/4}\;,
\end{equation}
where $\Gamma_\sigma$ is the spin fluctuation rate. Being a
classical result, the quantum-mechanical detailed-balancing property
is missing. Overall we thus write
\begin{equation}
  S_\sigma(\omega)=
  \frac{\Gamma_\sigma}{\omega^2+\Gamma_\sigma^2/4}\,
  s(\omega/T)\times \left\{
    \begin{array}{lr}
      1& \text{for}\;\; \omega\geq 0,\\
      {\rm e}^{\omega/T}&\text{for}\;\; \omega<0,
    \end{array}
  \right.
\end{equation}
where $s(x)$ is an even function normalized to $s(0)=1$. The axion
emission rate per unit mass, $\varepsilon_a=Q_a/\varrho$, therefore
is
\begin{equation}
  \varepsilon_a=
  \left(\frac{C_N}{2f_{a}}\right)^2\frac{T^4}{\pi^2m_N}\,F=
  3.0\times10^{37}\,\frac{\rm erg}{\rm g~s}\,
  C_N^2\left(\frac{\rm GeV}{f_{a}}\right)^2
  \left(\frac{T}{30\,{\rm MeV}}\right)^4 F\;,
\end{equation}
where
\begin{equation}
  F=\int_0^\infty dx\,\frac{x^4\,{\rm e}^{-x}}{4}\,
  \frac{\Gamma_\sigma/T}{x^2+(\Gamma_\sigma/2T)^2}\,s(x)\;.
\end{equation}
For $\Gamma_\sigma/T\ll 1$ (dilute medium) and assuming $s(x)=1$, we
find $F=\Gamma_\sigma/2T$.

A perturbative calculation, relevant for a dilute medium, and using
the OPE approximation, yields~\cite{GR:Brinkmann:1988vi,
GR:Raffelt:1993ix}
\begin{equation}
  \Gamma_\sigma^{\rm OPE}=4\pi^{1/2}\alpha_\pi^2\,
  \frac{n_BT^{1/2}}{m_N^{5/2}}=450\,{\rm MeV}\,
  \frac{\varrho}{3{\times}10^{14}\,{\rm g\,cm^{-3}}}\,
  \left(\frac{T}{30\,{\rm MeV}}\right)^{1/2}\;,
\end{equation}
where $\alpha_\pi=(f2m_N/m_\pi)^2/4\pi\approx15$ with $f\approx1.0$.
For soft energies, brems\-strah\-lung depends only on the on-shell
spin-dependent nucleon scattering rate. Based on measured nuclear
phase shifts it was argued that the OPE result was an overestimation
by about a factor of~4~\cite{GR:Hanhart:2000ae}.

Either way, $\Gamma_\sigma/T$ is not small compared to unity, but
also not very large. A possible range $1\lesssim
\Gamma_\sigma/T\lesssim10$ appears generous. With $s(x)=1$ this
would imply $F\approx0.5$ for $\Gamma_\sigma/T=1$, a maximum of
$F\approx 1.35$ near $\Gamma_\sigma/T=7$ and $F\approx 1.3$ for
$\Gamma_\sigma/T=10$. Of course, $s(x)$ probably decreases with $x$
or else the f-sum and higher sums of $S_\sigma(\omega)$ diverge. On
the basis of existing information one cannot do better than assume
$F$ to be a factor of order unity.

The SN~1987A limits are particularly interesting for hadronic axions
where the bounds on $\alpha_{aee}$ are moot. Therefore, we use
$C_p=-0.4$ and $C_n=0$. Initially the proton fraction is relatively
large so that we use $Y_p=0.3$ to scale the emission rate to the
proton density. With $F=1$ and $T=30\,{\rm MeV}$ we then find
$\varepsilon_a= 1.4\times10^{36}\,{\rm
  erg\,g^{-1}\,s^{-1}}$ so that~(\ref{eq:raffelt-SN1987A}) implies
\begin{equation}
  f_{a}\gtrsim4\times10^{8}\,{\rm GeV}\hbox{\qquad and\qquad}
  m_{a}\lesssim16\,{\rm meV}\;.
\end{equation}
Despite a lot of effort that has gone into understanding the axion
emission rate, these limits remain fairly rough estimates.

\section{Conclusions}
\label{sec:raffelt-conclusions}

\begin{figure}
  \sidecaption[t] 
  \includegraphics[width=0.5\textwidth]{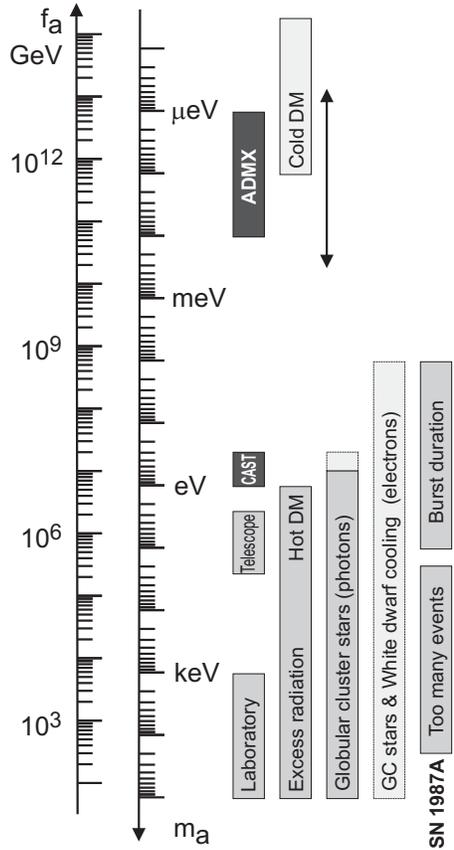}
  \caption{Summary of astrophysical and cosmological axion limits as
    discussed in the text. The black sensitivity bars indicate the search
    ranges of the CAST solar axion search and the ADMX search for galactic
    dark matter axions. Light-grey exclusion bars are very model
    dependent}\label{fig:raffelt-constraints}
\end{figure}

Astrophysics and cosmology provide the most restrictive limits on
the axion hypothesis as summarized in Fig.
\ref{fig:raffelt-constraints}.  Beginning with cosmology, a cold
axion population would emerge in the early universe that can make up
the dark matter, but the required axion mass involves many
uncertainties~\cite{GR:Sikivie:2006ni}. Galactic dark matter axions
will be searched by the ADMX experiment in the mass range
$m_{a}=1$--$100\,\umu{\rm eV}$~\cite{GR:Sikivie:1983ip,
GR:Bradley:2003kg, GR:Asztalos:2003px, GR:Duffy:2006aa}.

In addition, a population of hot axions is produced. Before
confinement, the relevant processes involve quarks and
gluons~\cite{GR:Turner:1986tb, GR:Masso:2002np}. Later the most
generic process is $\pi+\pi\leftrightarrow\pi+a$
\cite{GR:Chang:1993gm}. Axions decouple after the QCD epoch if
$f_{a}\lesssim3\times10^7\,{\rm GeV}$ ($m_{a}\gtrsim0.2\,{\rm
  eV}$).  Some of these hot dark matter axions would be trapped in galaxies
and galaxy clusters. An unsuccesful search for a decay
line~\cite{GR:Bershady:1990sw, GR:Ressell:1991zv, Grin:2006aw}
provides direct limits on a range of axion masses marked
``Telescope'' in Fig. \ref{fig:raffelt-constraints}.  Moreover, the
usual structure-formation arguments provide the hot-dark matter
limits \cite{GR:Hannestad:2003ye, GR:Hannestad:2005df}. Axions decay
on a cosmic time scale for $m_{a}\gtrsim20\,{\rm eV}$. The decay
photons would cause a variety of observable
consequences~\cite{GR:Masso:1997ru}, seamlessly connecting with the
hot dark matter limit so that cosmology alone rules out axions in
the entire mass range $m_{a}>1\,{\rm eV}$.

Figure \ref{fig:raffelt-constraints} also shows the
stellar-evolution limits discussed in this chapter, notably the
globular-cluster limit on the axion-photon coupling. The
globular-cluster and white-dwarf limits for DFSZ axions with
$\cos^2\beta=1$ are shown as a light-grey exclusion bar. Even in the
DFSZ model, the axion-electron coupling could be accidentally small
and at tree level it is entirely absent for hadronic axions.
Therefore, these limits are far less generic than those based on the
axion-photon or axion-nucleon interaction.

The requirement that the neutrino signal of SN~1987A was not
excessively shortened by axion losses pushes the limits down to
$m_{a}\lesssim 10\,{\rm meV}$. However, this limit involves many
uncertainties that are difficult to quantify so that it is somewhat
schematic. The CAST search for solar
axions~\cite{GR:Andriamonje:2004hi} covers new territory in the
parameter plane of $m_{a}$ and $g_{a\gamma\gamma}$, but a signal
would represent a conflict with the SN~1987A limit. While this limit
certainly suggests that axions more plausibly have masses relevant
for cold dark matter, a single argument, measurement or observation
is never conclusive.

In the DFSZ model, the limits from white-dwarf cooling based on the
axion-electron interaction and those from SN~1987A from the
axion-nucleon interaction are quite similar. Therefore, axion
emission could still play an important role as an energy-loss
channel of both SNe and white dwarfs and for other evolved stars,
e.g.\ asymptotic giant stars.

In summary, axions provide a show-case example for the fascinating
interplay between astrophysics, cosmology and particle physics to
solve some of the deepest mysteries at the interface between inner
space and outer space.

\section*{Acknowlegments}
\label{subsec:raffelt-acknowlegments}

Partial support by the Deutsche Forschungsgemeinschaft under grant
SFB 375 and the European Union, contract RII3-CT-2004-506222 (ILIAS
project), is acknowledged.



\begin{thebibliography}{99.}

\bibitem{GR:Gamow:1940}
  G.~Gamow and M.~Schoenberg,
  ``The possible role of neutrinos in stellar evolution,''
  Phys.\ Rev.\ {58}, 1117 (1940).

\bibitem{GR:Gamow:1941}
  G.~Gamow and M.~Schoenberg,
  ``Neutrino theory of stellar collapse,''
  Phys.\ Rev.\ {59}, 539 (1941).

\bibitem{GR:Bernstein:1963qh}
  J.~Bernstein, M.~Ruderman and G.~Feinberg,
  ``Electromagnetic properties of the neutrino,''
  Phys.\ Rev.\  {132}, 1227 (1963).

\bibitem{GR:Stothers:1970ap}
  R.~B.~Stothers,
  ``Astrophysical determination of the coupling constant for the
  electron-neutrino weak interaction,''
  Phys.\ Rev.\ Lett.\ {24}, 538 (1970).

\bibitem{GR:Sato:1975vy}
  K.~Sato and H.~Sato,
  ``Higgs meson emission from a star and a constraint on its mass,''
  Prog.\ Theor.\ Phys.\ {54}, 1564 (1975).

\bibitem{GR:Dicus:fp}
  D.~A.~Dicus, E.~W.~Kolb, V.~L.~Teplitz and R.~V.~Wagoner,
  ``Astrophysical bounds on the masses of axions and Higgs
  particles,''
  Phys.\ Rev.\ D {18}, 1829 (1978).

\bibitem{GR:Vysotsky:1978dc}
  M.~I.~Vysotsky, Y.~B.~Zeldovich, M.~Y.~Khlopov and V.~M.~Chechetkin,
  ``Some astrophysical limitations on the axion mass,''
  Pisma Zh.\ Eksp.\ Teor.\ Fiz.\ {27}, 533 (1978)
  [JEPT Lett.\ {27}, 502 (1978)].

\bibitem{GR:Turner:1989vc}
  M.~S.~Turner,
  ``Windows on the axion,''
  Phys.\ Rept.\ {197}, 67 (1990).

\bibitem{GR:Raffelt:1990yz}
  G.~G.~Raffelt,
  ``Astrophysical methods to constrain axions and other novel particle
  phenomena,''
  Phys.\ Rept.\ {198}, 1 (1990).

\bibitem{GR:Raffelt:1996wa}
  G.~G.~Raffelt,
  {\it Stars as laboratories for fundamental physics\/}
  (University of Chicago Press, 1996).

\bibitem{GR:Raffelt:1999tx}
  G.~G.~Raffelt,
  ``Particle physics from stars,''
  Ann.\ Rev.\ Nucl.\ Part.\ Sci.\  {49}, 163 (1999)
  [hep-ph/9903472].

\bibitem{GR:Yao:2006px}
  W.~M.~Yao {et al.}  (Particle Data Group),
  ``Review of particle physics,''
  J.\ Phys.\ G {33}, 1 (2006).

\bibitem{GR:Georgi:1986df}
  H.~Georgi, D.~B.~Kaplan and L.~Randall,
  ``Manifesting the invisible axion at low energies,''
  Phys.\ Lett.\ B {169}, 73 (1986).

\bibitem{GR:Kamionkowski:1992mf}
  M.~Kamionkowski and J.~March-Russell,
  ``Planck scale physics and the Peccei-Quinn mechanism,''
  Phys.\ Lett.\ B {282}, 137 (1992)
  [hep-th/9202003].

\bibitem{GR:Barr:1992qq}
  S.~M.~Barr and D.~Seckel,
  ``Planck scale corrections to axion models,''
  Phys.\ Rev.\ D {46}, 539 (1992).

\bibitem{GR:Gasser:1982ap}
  J.~Gasser and H.~Leutwyler,
  ``Quark masses,''
  Phys.\ Rept.\ {87}, 77 (1982).

\bibitem{GR:Leutwyler:1996qg}
  H.~Leutwyler,
  ``The ratios of the light quark masses,''
  Phys.\ Lett.\ B {378}, 313 (1996)
  [hep-ph/9602366].

\bibitem{GR:Kim:1979if}
  J.~E.~Kim,
  ``Weak interaction singlet and strong CP invariance,''
  Phys.\ Rev.\ Lett.\  {43}, 103 (1979).

\bibitem{GR:Shifman:1979if}
  M.~A.~Shifman, A.~I.~Vainshtein and V.~I.~Zakharov,
  ``Can confinement ensure natural CP invariance of strong
  interactions?,''
  Nucl.\ Phys.\ B {166}, 493 (1980).

\bibitem{GR:Zhitnitsky:1980tq}
  A.~R.~Zhitnitsky,
  ``On possible suppression of the axion hadron interactions,''
  Sov.\ J.\ Nucl.\ Phys.\ {31}, 260 (1980)
  [Yad.\ Fiz.\ {31}, 497 (1980)].

\bibitem{GR:Dine:1981rt}
  M.~Dine, W.~Fischler and M.~Srednicki,
  ``A simple solution to the strong CP problem with a harmless
  axion,''
  Phys.\ Lett.\ B {104}, 199 (1981).

\bibitem{GR:Cheng:1995fd}
  S.~L.~Cheng, C.~Q.~Geng and W.~T.~Ni,
  ``Axion-photon couplings in invisible axion models,''
  Phys.\ Rev.\ D {52}, 3132 (1995)
  [hep-ph/9506295].

\bibitem{GR:Raffelt:1987yt}
  G.~Raffelt and D.~Seckel,
  ``Bounds on exotic particle interactions from SN 1987A''
  Phys.\ Rev.\ Lett.\  {60}, 1793 (1988).

\bibitem{GR:Carena:1988kr}
  M.~Carena and R.~D.~Peccei,
  ``The effective Lagrangian for axion emission from SN 1987A,''
  Phys.\ Rev.\ D {40}, 652 (1989).

\bibitem{GR:Alexakhin:2006vx}
  V.~Y.~Alexakhin {et al.} (COMPASS Collaboration),
  ``The deuteron spin-dependent structure function $g_1^d$ and
  its first moment,''
  hep-ex/0609038.

\bibitem{GR:Airapetian:2006vy}
  A.~Airapetian (HERMES Collaboration),
  ``Precise determination of the spin structure function
  $g_1$ of the proton, deuteron and neutron,''
  hep-ex/0609039.

\bibitem{GR:Ellis:1995de}
  J.~R.~Ellis and M.~Karliner,
  ``The strange spin of the nucleon,''
  in: {\it The Spin Structure of the Nucleon:
  International School of Nucleon Structure\/}
  (3--10 August 1995, Erice, Italy), edited by
  B.~Frois, V.~W.~Hughes and N.~De Groot
  (World Scientific, Singapore, 1997)
  [hep-ph/9601280].

\bibitem{GR:Primakoff}
  H.~Primakoff,
  ``Photo-production of neutral mesons in nuclear electric fields
  and the mean life of the neutral meson,''
  Phys.\ Rev.\ {81}, 899 (1951).

\bibitem{GR:Raffelt:1985nk}
  G.~G.~Raffelt,
  ``Astrophysical axion bounds diminished by screening effects,''
  Phys.\ Rev.\ D {33}, 897 (1986).

\bibitem{GR:Altherr:1993zd}
  T.~Altherr, E.~Petitgirard and T.~del R\'\i o Gaztelurrutia,
  ``Axion emission from red giants and white dwarfs,''
  Astropart.\ Phys.\ {2}, 175 (1994)
  [hep-ph/9310304].

\bibitem{GR:Raffelt:1987np}
  G.~G.~Raffelt,
  ``Plasmon decay into low mass bosons in stars,''
  Phys.\ Rev.\ D {37}, 1356 (1988).

\bibitem{GR:Zavattini:2005tm}
  E.~Zavattini {et al.}  (PVLAS Collaboration),
  ``Experimental observation of optical rotation generated
  in vacuum by a magnetic field,''
  Phys.\ Rev.\ Lett.\  {96}, 110406 (2006)
  [hep-ex/0507107].

\bibitem{GR:Cameron:1993mr}
  R.~Cameron {et al.},
  ``Search for nearly massless, weakly coupled particles by
  optical techniques,''
  Phys.\ Rev.\ D {47}, 3707 (1993).

\bibitem{GR:Schlattl:1998fz}
  H.~Schlattl, A.~Weiss and G.~Raffelt,
  ``Helioseismological constraint on solar axion emission,''
  Astropart.\ Phys.\  {10}, 353 (1999)
  [hep-ph/9807476].

\bibitem{GR:Bahcall:2004pz}
  J.~N.~Bahcall, A.~M.~Serenelli and S.~Basu,
  ``New solar opacities, abundances, helioseismology,
  and neutrino fluxes,''
  Astrophys.\ J.\  {621}, L85 (2005)
  [astro-ph/0412440].

\bibitem{GR:Ahmad:2002jz}
  Q.~R.~Ahmad {et al.}  (SNO Collaboration),
  ``Direct evidence for neutrino flavor transformation from
  neutral-current interactions in the Sudbury Neutrino Observatory,''
  Phys.\ Rev.\ Lett.\  {89}, 011301 (2002)
  [nucl-ex/0204008].

\bibitem{GR:Aharmim:2005gt}
  B.~Aharmim {et al.}  (SNO Collaboration),
  ``Electron energy spectra, fluxes, and day-night asymmetries of B-8
   solar neutrinos from the 391-day salt phase SNO data set,''
  Phys.\ Rev.\ C {72}, 055502 (2005)
  [nucl-ex/0502021].

\bibitem{GR:Sikivie:1983ip}
  P.~Sikivie,
  ``Experimental tests of the invisible axion,''
  Phys.\ Rev.\ Lett.\  {51}, 1415 (1983);
  (E) ibid.\ {52}, 695 (1984).

\bibitem{GR:Raffelt:1987im}
  G.~Raffelt and L.~Stodolsky,
  ``Mixing of the photon with low mass particles,''
  Phys.\ Rev.\ D {37}, 1237 (1988).

\bibitem{GR:Lazarus:1992ry}
  D.~M.~Lazarus, G.~C.~Smith, R.~Cameron, A.~C.~Melissinos,
  G.~Ruoso, Y.~K.~Semertzidis and F.~A.~Nezrick,
  ``A search for solar axions,''
  Phys.\ Rev.\ Lett.\  {69}, 2333 (1992).

\bibitem{GR:Moriyama:1998kd}
  S.~Moriyama, M.~Minowa, T.~Namba, Y.~Inoue, Y.~Takasu
  and A.~Yamamoto,
  ``Direct search for solar axions by using strong magnetic field and
  x-ray detectors,''
  Phys.\ Lett.\ B {434}, 147 (1998)
  [hep-ex/9805026].

\bibitem{GR:Inoue:2002qy}
  Y.~Inoue, T.~Namba, S.~Moriyama, M.~Minowa, Y.~Takasu,
  T.~Horiuchi and A.~Yamamoto,
  ``Search for sub-electronvolt solar axions using coherent
  conversion of axions into photons in magnetic field and
  gas helium,''
  Phys.\ Lett.\ B {536}, 18 (2002)
  [astro-ph/0204388].

\bibitem{GR:Avignone:1997th}
  F.~T.~Avignone {et al.} (SOLAX Collaboration),
  ``Experimental search for solar axions via coherent
  Primakoff conversion  in a germanium spectrometer,''
  Phys.\ Rev.\ Lett.\ {81}, 5068 (1998)
  [astro-ph/9708008].

\bibitem{GR:Morales:2001we}
  A.~Morales {et al.} (COSME Collaboration),
  ``Particle dark matter and solar axion searches with a small
  germanium detector at the Canfranc underground laboratory,''
  Astropart.\ Phys.\ {16}, 325 (2002)
  [hep-ex/0101037].

\bibitem{GR:Bernabei:ny}
  R.~Bernabei {et al.},
  ``Search for solar axions by Primakoff effect in NaI crystals,''
  Phys.\ Lett.\ B {515}, 6 (2001).

\bibitem{GR:Andriamonje:2004hi}
  K.~Zioutas {et al.}  (CAST Collaboration),
  ``First results from the CERN axion solar telescope (CAST),''
  Phys.\ Rev.\ Lett.\ {94}, 121301 (2005)
  [hep-ex/0411033].

\bibitem{GR:vanBibber:1988ge}
  K.~van Bibber, P.~M.~McIntyre, D.~E.~Morris and G.~G.~Raffelt,
  ``A practical laboratory detector for solar axions,''
  Phys.\ Rev.\ D {39}, 2089 (1989).

\bibitem{GR:Raffelt:1988rx}
  G.~G.~Raffelt and G.~D.~Starkman,
  ``Stellar energy transfer by keV mass scalars,''
  Phys.\ Rev.\ D {40}, 942 (1989).

\bibitem{GR:Buzzoni:1983}
  A.~Buzzoni, F.~Fusi Pecci, R.~Buonanno and C.~E.~Corsi
  ``Helium abundance in globular clusters: the R-method''
  Astron.\ Astrophys.\ {128}, 94 (1983).

\bibitem{GR:Buonanno:1986}
  R.~Buonanno, A.~Buzzoni, C.~E.~Corsi, F.~Fusi Pecci and A.~R.~Sandage,
  ``High precision photometry of 10000 stars in M3,''
  Mem.\ Soc.\ Astron.\ Ital.\ {57}, 391 (1986).

\bibitem{GR:Renzini:1988}
  A.~Renzini and F.~Fusi Pecci,
  ``Tests of evolutionary sequences using color-magnitude diagrams of globular
  clusters,''
  Annu.\ Rev.\ Astron.\ Astrophys.\ {26}, 199 (1988).

\bibitem{GR:Raffelt:1989xu}
  G.~G.~Raffelt,
  ``Core mass at the helium flash from observations and a new
  bound on neutrino electromagnetic properties,''
  Astrophys.\ J.\  {365}, 559 (1990).

\bibitem{GR:Raffelt:1990pj}
  G.~G.~Raffelt,
  ``New bound on neutrino dipole moments from globular cluster stars,''
  Phys.\ Rev.\ Lett.\  {64}, 2856 (1990).

\bibitem{GR:Raffelt:1994ry}
  G.~Raffelt and A.~Weiss,
  ``Red giant bound on the axion-electron coupling revisited,''
  Phys.\ Rev.\ D {51}, 1495 (1995)
  [hep-ph/9410205].

\bibitem{GR:Catelan:1995ba}
  M.~Catelan, J.~A.~de Freitas Pacheco and J.~E.~Horvath,
  ``The helium-core mass at the helium flash in low-mass
  red giant stars: Observations and theory,''
  Astrophys.\ J.\ {461}, 231 (1996)
  [astro-ph/9509062].

\bibitem{GR:Dominguez:1999gg}
  I.~Dom{\'\i}nguez, O.~Straniero and J.~Isern,
  ``Asymptotic giant branch stars as astroparticle laboratories,''
  Mon. Not. R. Astron. Soc. {306}, L1 (1999)
  [astro-ph/9905033].

\bibitem{GR:Raffelt:1985nj}
  G.~G.~Raffelt,
  ``Axion constraints from white dwarf cooling times,''
  Phys.\ Lett.\ B {166}, 402 (1986).

\bibitem{GR:Wang:1992gc}
  J.~Wang,
  ``Constraints of axions from white dwarf cooling,''
  Mod.\ Phys.\ Lett.\ A {7}, 1497 (1992).

\bibitem{GR:Blinnikov:1994}
  S.~I.~Blinnikov and N.~V.~Dunina-Barkovskaya,
  ``The cooling of hot white dwarfs: A theory with non-standard
  weak interactions and a comparison with observations,''
  Mon.\ Not.\ R.\ Astron.\ Soc.\ {266}, 289 (1994).

\bibitem{GR:Isern:1992}
  J.~Isern, M.~Hernanz and E.~Garc{\'\i}a-Berro,
  ``Axion cooling of white dwarfs,''
  Astrophys.\ J.\ {392}, L23 (1992).

\bibitem{GR:Corsico:2001be}
  A.~H.~C\'orsico, O.~G.~Benvenuto, L.~G.~Althaus, J.~Isern
  and E.~Garc{\'\i}a-Berro,
  ``The potential of the variable DA white dwarf G117-B15A as a tool
  for fundamental physics,''
  New Astron.\ {6}, 197 (2001)
  [astro-ph/0104103].

\bibitem{GR:Isern:2003xj}
  J.~Isern and E.~Garc{\'\i}a-Berro,
  ``White dwarf stars as particle physics laboratories,''
  Nucl.\ Phys.\ Proc.\ Suppl.\ {114}, 107 (2003).

\bibitem{GR:Koshiba:1992yb}
  M.~Koshiba,
  ``Observational neutrino astrophysics,''
  Phys.\ Rept.\  {220}, 229 (1992).

\bibitem{GR:Burrows:2000mk}
  A.~Burrows,
  ``Supernova explosions in the universe,''
  Nature {403}, 727 (2000).

\bibitem{GR:Woosley:2006ie}
  S.~Woosley and T.~Janka,
  ``The physics of core-collapse supernovae,''
  Nature Physics {1}, 147 (2005)
  [astro-ph/0601261].

\bibitem{GR:Ellis:1987pk}
  J.~R.~Ellis and K.~A.~Olive,
  ``Constraints on light particles from supernova 1987A''
  Phys.\ Lett.\ B {193}, 525 (1987).

\bibitem{GR:Turner:1987by}
  M.~S.~Turner,
  ``Axions from SN~1987A,''
  Phys.\ Rev.\ Lett.\  {60}, 1797 (1988).

\bibitem{GR:Mayle:1987as}
  R.~Mayle, J.~R.~Wilson, J.~R.~Ellis, K.~A.~Olive,
  D.~N.~Schramm and G.~Steigman,
  ``Constraints on axions from SN 1987A''
  Phys.\ Lett.\ B {203}, 188 (1988).

\bibitem{GR:Mayle:1989yx}
  R.~Mayle, J.~R.~Wilson, J.~R.~Ellis, K.~A.~Olive,
  D.~N.~Schramm and G.~Steigman,
  ``Updated constraints on axions from SN~1987A,''
  Phys.\ Lett.\ B {219}, 515 (1989).

\bibitem{GR:Brinkmann:1988vi}
  R.~P.~Brinkmann and M.~S.~Turner,
  ``Numerical rates for nucleon-nucleon axion bremsstrahlung,''
  Phys.\ Rev.\ D {38}, 2338 (1988).

\bibitem{GR:Burrows:1988ah}
  A.~Burrows, M.~S.~Turner and R.~P.~Brinkmann,
  ``Axions and SN~1987A,''
  Phys.\ Rev.\ D {39}, 1020 (1989).

\bibitem{GR:Burrows:1990pk}
  A.~Burrows, M.~T.~Ressell and M.~S.~Turner,
  ``Axions and SN~1987A: Axion trapping,''
  Phys.\ Rev.\ D {42}, 3297 (1990).

\bibitem{GR:Janka:1995ir}
  H.~T.~Janka, W.~Keil, G.~Raffelt and D.~Seckel,
  ``Nucleon spin fluctuations and the supernova
  emission of neutrinos and axions,''
  Phys.\ Rev.\ Lett.\ {76}, 2621 (1996)
  [astro-ph/9507023].

\bibitem{GR:Keil:1996ju}
  W.~Keil, H.~T.~Janka, D.~N.~Schramm, G.~Sigl,
  M.~S.~Turner and J.~R.~Ellis,
  ``A fresh look at axions and SN 1987A,''
  Phys.\ Rev.\ D {56}, 2419 (1997)
  [astro-ph/9612222].

\bibitem{GR:Hanhart:2000ae}
  C.~Hanhart, D.~R.~Phillips and S.~Reddy,
  ``Neutrino and axion emissivities of neutron stars
  from nucleon nucleon scattering data,''
  Phys.\ Lett.\ B {499}, 9 (2001)
  [astro-ph/0003445].

\bibitem{GR:Engel:1990zd}
  J.~Engel, D.~Seckel and A.~C.~Hayes,
  ``Emission and detectability of hadronic axions
  from SN~1987A,''
  Phys.\ Rev.\ Lett.\  {65}, 960 (1990).

\bibitem{GR:Hanhart:2001fx}
  C.~Hanhart, J.~A.~Pons, D.~R.~Phillips and S.~Reddy,
  ``The likelihood of GODs' existence: Improving the SN 1987A
  constraint on the size of large compact dimensions,''
  Phys.\ Lett.\ B {509}, 1 (2001)
  [astro-ph/0102063].

\bibitem{GR:Raffelt:1991pw}
  G.~Raffelt and D.~Seckel,
  ``Multiple scattering suppression of the bremsstrahlung emission
  of neutrinos and axions in supernovae,''
  Phys.\ Rev.\ Lett.\  {67}, 2605 (1991).

\bibitem{GR:Raffelt:1993ix}
  G.~Raffelt and D.~Seckel,
  ``A selfconsistent approach to neutral current processes in
  supernova cores,''
  Phys.\ Rev.\ D {52}, 1780 (1995)
  [astro-ph/9312019].

\bibitem{GR:Raffelt:1996za}
  G.~Raffelt, D.~Seckel and G.~Sigl,
  ``Supernova neutrino scattering rates reduced by nucleon
  spin fluctuations: Perturbative limit,''
  Phys.\ Rev.\ D {54}, 2784 (1996)
  [astro-ph/9603044].

\bibitem{GR:Raffelt:1996di}
  G.~Raffelt and T.~Strobel,
  ``Reduction of weak interaction rates in neutron stars by nucleon
  spin fluctuations: Degenerate case,''
  Phys.\ Rev.\ D {55}, 523 (1997)
  [astro-ph/9610193].

\bibitem{GR:Sigl:1995ac}
  G.~Sigl,
  ``Weak interactions in supernova cores and saturation of
  nucleon spin fluctuations,''
  Phys.\ Rev.\ Lett.\  {76}, 2625 (1996)
  [astro-ph/9508046].

\bibitem{GR:Raffelt:1998pa}
  G.~Raffelt and G.~Sigl,
  ``Numerical toy-model calculation of the nucleon spin
  autocorrelation function in a supernova core,''
  Phys.\ Rev.\ D {60}, 023001 (1999)
  [hep-ph/9808476].

\bibitem{GR:Yamada:1999va}
  S.~Yamada,
  ``Reduction of neutrino nucleon scattering rate by nucleon nucleon
  collisions,''
  Nucl.\ Phys.\ A {662}, 219 (2000)
  [astro-ph/9907045].

\bibitem{GR:Sedrakian:2000kc}
  A.~Sedrakian and A.~E.~L.~Dieperink,
  ``Coherent neutrino radiation in supernovae at two loops,''
  Phys.\ Rev.\ D {62}, 083002 (2000)
  [astro-ph/0002228].

\bibitem{GR:vanDalen:2003zw}
  E.~N.~E.~van Dalen, A.~E.~L.~Dieperink and J.~A.~Tjon,
  ``Neutrino emission in neutron stars,''
  Phys.\ Rev.\ C {67}, 065807 (2003)
  [nucl-th/0303037].

\bibitem{GR:Sikivie:2006ni}
  P.~Sikivie,
  ``Axion cosmology,''
  astro-ph/0610440.

\bibitem{GR:Bradley:2003kg}
  R.~Bradley {et al.},
  ``Microwave cavity searches for dark-matter axions,''
  Rev.\ Mod.\ Phys.\  {75}, 777 (2003).

\bibitem{GR:Asztalos:2003px}
  S.~J.~Asztalos {et al.},
  ``An improved RF cavity search for halo axions,''
  Phys.\ Rev.\ D {69}, 011101 (2004)
  [astro-ph/0310042].

\bibitem{GR:Duffy:2006aa}
  L.~D.~Duffy {et al.},
  ``A high resolution search for dark-matter axions,''
  Phys.\ Rev.\ D {74}, 012006 (2006)
  [astro-ph/0603108].

\bibitem{GR:Turner:1986tb}
  M.~S.~Turner,
  ``Thermal production of not so invisible axions in the
  early universe,''
  Phys.\ Rev.\ Lett.\  {59}, 2489 (1987);
  (E) ibid.\ {60}, 1101 (1988).

\bibitem{GR:Masso:2002np}
  E.~Mass\'o, F.~Rota and G.~Zsembinszki,
  ``On axion thermalization in the early universe,''
  Phys.\ Rev.\ D {66}, 023004 (2002)
  [hep-ph/0203221].

\bibitem{GR:Chang:1993gm}
  S.~Chang and K.~Choi,
  ``Hadronic axion window and the big bang nucleosynthesis,''
  Phys.\ Lett.\ B {316}, 51 (1993)
  [hep-ph/9306216].

\bibitem{GR:Bershady:1990sw}
  M.~A.~Bershady, M.~T.~Ressell and M.~S.~Turner,
  ``Telescope search for multi-eV axions,''
  Phys.\ Rev.\ Lett.\  {66}, 1398 (1991).

\bibitem{GR:Ressell:1991zv}
  M.~T.~Ressell,
  ``Limits to the radiative decay of the axion,''
  Phys.\ Rev.\ D {44}, 3001 (1991).

\bibitem{Grin:2006aw}
  D.~Grin, G.~Covone, J.~P.~Kneib, M.~Kamionkowski, A.~Blain
  and E.~Jullo,
  ``A telescope search for decaying relic axions,''
  astro-ph/0611502.

\bibitem{GR:Hannestad:2003ye}
  S.~Hannestad and G.~Raffelt,
  ``Cosmological mass limits on neutrinos, axions, and other light
  particles,''
  JCAP {0404}, 008 (2004)
  [hep-ph/0312154].

\bibitem{GR:Hannestad:2005df}
  S.~Hannestad, A.~Mirizzi and G.~Raffelt,
  ``New cosmological mass limit on thermal relic axions,''
  JCAP {0507}, 002 (2005)
  [hep-ph/0504059].

\bibitem{GR:Masso:1997ru}
  E.~Mass\'o and R.~Toldra,
  ``New constraints on a light spinless particle coupled to photons,''
  Phys.\ Rev.\ D {55}, 7967 (1997)
  [hep-ph/9702275].

\end{thebibliography}
\end{document}